%% file: qcdnumxx.tex
\documentclass[12pt,twoside,a4paper]{article}

  \usepackage{epstopdf}
  \usepackage{amssymb}
  \usepackage{amsmath}
  \usepackage{minitoc}
  \usepackage{tikz}
  \usetikzlibrary{calc}
  \usepackage[framemethod=tikz]{mdframed}
  \usepackage[margin=10pt,font=footnotesize,labelfont=bf,
               labelsep=endash]{caption}
               
  \input layout.tex
  \mtcsetfont{secttoc}{subsection}{\normalfont}
  
  \input newcommands.tex

\title{
  \Large \bf A \cxx\ interface to QCDNUM
}

\author{
  V.~Bertone\thanks{email valerio.bertone@cern.ch}\ , 
  M.~Botje\thanks{email m.botje@nikhef.nl}\\
  NIKHEF, Science Park, Amsterdam, the Netherlands
}

\date{\today}

\begin{document}

\maketitle 
 
\begin{abstract}
\noindent
In this document we report on the recent development of a \cxx\
interface to the \fortran-based evolution program \qcdnum. A
short description of the interface is given with a few basic
examples of its usage.
\end{abstract}

\section{Introduction \label{se:Introduction}}

The \qcdnum\ program~\cite{ref:qcdnum} numerically solves the DGLAP
evolution equations in perturbative QCD up to next-to-next-to-leading
order (NNLO) in the strong coupling \as\ for unpolarised parton distribution
functions~(PDFs) and up to next-to-leading order (NLO) for 
polarised PDFs and fragmentation
functions. \Qcdnum~also provides the users with a set
of flexible tools for the fast computation of customised evolutions
and physical observables. Based on these tools, and
included in the \qcdnum\ distribution, are packages for
the computation
of the unpolarised deep-inelastic-scattering structure functions
in the zero-mass scheme and heavy-flavour contributions to these
structure functions in the massive scheme up to
order $\as^2$. 

Thanks to its capabilities, \qcdnum~has been employed in a large
number of phenomenological studies, see for example the recent
results of Refs.~\cite{Vafaee:2017jiv,Andreev:2017vxu,Hautmann:2017fcj}.
\Qcdnum\ is currently interfaced to a number of codes routinely used for the
analysis and the interpretation of experimental data, such as {\tt
  xFitter}~\cite{Alekhin:2014irh} and {\tt Alpos}~\cite{ref:Alpos}.

The \qcdnum\ code is written in~\fortranzz. This language
usually guarantees an excellent performance, thanks to its
simplicity and the static management of memory. Indeed, speed is one of
the strongest points of \qcdnum\ that mostly derives from the use of
\fortranzz\ but also from an optimal memory-access
strategy made possible by an in-house dynamic memory manager. However,
it is a fact that in today's large-scale HEP codes \fortran\ is abandoned
in favour of object-oriented languages like \cxx\ with its built-in 
dynamic memory allocation and support for modularity. It is therefore
important that standard tools, such as
PDF evolution codes, are not only available in \fortran\ but also
in \cxx.

Instead of re-writing \qcdnum\ entirely
in \cxx\ (a huge effort which might yield degraded performance in terms
of speed) we have decided to provide
an interface that allows users to call the existing \qcdnum\ routines from
a \cxx\ program. In this note  we present the~\cxx\ interface that
has been made available in the beta release 
\mbox{17-01/14} which can be downloaded from the 
web-site given in Ref.~\cite{ref:QCDNUMweb}.
The interface is also documented in the \qcdnum\ write-up. 

\section{The \cxx\ interface \label{se:interface}}

Since release 17-01/10, the \qcdnum\ library can be built using
\autotools~\cite{ref:autotools} which is a standard suite of tools
to make source code packages portable on different
platforms. \Autotools\ provides some support
to facilitate the simultaneous use of multiple languages, such as the
automatic selection of the linker and its flags and macros to convert
names of \cxx\ identifiers into the \fortran\ format. 

This allowed us to write \cxx\ wrappers to the original \fortran\ 
routines so that, from version
\mbox{17-01/14} onward,\footnote{At present, wrappers are
available for the out-of-the-box routines listed in Table~2 of the
\qcdnum\ write-up, except that the steering by data cards is only 
possible in \fortran. Also available are wrappers
for the \zmstf\ and \hqstf\ structure function packages. 
Those for the \qcdnum\ toolbox (Table~4)
are under development and their validation will be announced
 on the \qcdnum\ web site and mailing list.} 
it is possible to call \qcdnum\ from a \cxx\ code, provided that the
library is built with \autotools, and not locally with 
the~\xtt{makelib} script.
\bcxx
\begin{verbatim}
      bash> cd qcdnum-17-01-14
      bash> ./configure --prefix=<installation-dir>
      bash>  make
      bash>  make install
\end{verbatim}
\ecxx
The \qcdnum\ includes, library and the \xtt{qcdnum-config} script can then
be found in, respectively, the sub-directories
\xtt{/include}, \xtt{/lib} and \xtt{/bin} of \xtt{<installation-dir>}.
To make these accessible from everywhere, care should be taken to 
properly set-up the search paths, for instance on OS-X,
\bcxx
\begin{verbatim}
      bash> export PATH=$PATH:<installation-dir>/bin
      bash> export LD_LIBRARY_PATH=$LD_LIBRARY_PATH:<installation-dir>/lib
\end{verbatim}
\ecxx
Now one can run test programs that are stored in the 
\xtt{testjobsCxx} folder.
\bcxx 
\begin{verbatim}
      bash> cd qcdnum-17-01-14/run
      bash> ./runtest exampleCxx         (without extension .cc)
\end{verbatim}      
\ecxx
The program \xtt{exampleCxx.cc} evolves the set of unpolarised 
parton densities at NNLO in the variable flavour number scheme (VFNS)
and prints the value of the charm sea~$2x\cbar$ at some $x$ and \ms, as well
as the value of \asmz. The output should look like this.
\bcxx
\begin{verbatim}
      x, q, CharmSea = 1.0000e-03  1.0000e+03  1.8708e+00
      as(mz2)        = 1.1807e-01
\end{verbatim}
\ecxx
%
\section{Example code\label{se:example}}

To illustrate how the interface is used in practice, we show
in \Fi{fig:cxxlisting} 
%
\begin{figure}
\bcxx
\begin{verbatim}
  ...
#include "QCDNUM/QCDNUM.h"
  ...

double func(int* ipdf, double* x) {
  int i = *ipdf;
  double xb = *x; 
  double f = 0;
  if(i ==  0) f = xglu(xb);
  if(i ==  1) f = xdnv(xb);
  if(i ==  2) f = xupv(xb);
  ...  
  return f;
  }

int main() {
  ...
  double def[] =                          //input flavour composition
  // tb  bb  cb  sb  ub  db   g   d   u   s   c   b   t
    { 0., 0., 0., 0., 0.,-1., 0., 1., 0., 0., 0., 0., 0.,   // 1=dval
      0., 0., 0., 0.,-1., 0., 0., 0., 1., 0., 0., 0., 0.,   // 2=uval
      0., 0., 0.,-1., 0., 0., 0., 0., 0., 1., 0., 0., 0.,   // 3=sval
      0., 0., 0., 0., 0., 1., 0., 0., 0., 0., 0., 0., 0.,   // 4=dbar
      0., 0., 0., 0., 1., 0., 0., 0., 0., 0., 0., 0., 0.,   // 5=ubar
      0., 0., 0., 1., 0., 0., 0., 0., 0., 0., 0., 0., 0.,   // 6=sbar
      ...                                           // more not shown
      0., 0., 0., 0., 0., 0., 0., 0., 0., 0., 0., 0., 0.};  //12=zero
  ...
  QCDNUM::qcinit(lun,outfile);                   //initialise
  QCDNUM::gxmake(xmin,iwt,ng,nxin,nx,iosp);      //x-grid
  QCDNUM::gqmake(qq,wt,2,nqin,nq);               //mu2-grid
  QCDNUM::setcbt(nfin,iqc,iqb,999);              //set VFNS thresholds
  QCDNUM::fillwt(ityp,id1,id2,nw);               //compute weights
  QCDNUM::setord(iord);                          //LO, NLO, NNLO
  QCDNUM::setalf(as0,r20);                       //input alphas
  QCDNUM::evolfg(ityp,func,def,iq0,eps);         //evolve all pdf's
  QCDNUM::allfxq(ityp,x,q,pdf,0,1);              //interpolate all pdf's
  
  double csea = 2 * pdf[2];                      //charm sea at x,mu2
  double asmz = QCDNUM::asfunc(qmz2,nfout,ierr); //alphas(mz2)
  
  cout << scientific << setprecision(4);
  cout << "x, q, CharmSea = " << x << "  " << q << "  " << csea << endl;
  cout << "as(mz2)        = " << asmz << endl;
  
  return 0;
  }      
\end{verbatim}
\ecxx
\caption{\footnotesize A \qcdnum\ example program in \cxx.
Only the relevant parts of the code are shown.}
\label{fig:cxxlisting}
\end{figure}
%
a listing of the~\xtt{exampleCxx} program. For the sake of clarity
we have omitted some include directives, type declarations and
variable initialisations; for a full listing the reader is referred
to the code stored in the \xtt{testjobsCxx} directory of the \qcdnum\
distribution. We will not describe here the function calls in
detail (see the \qcdnum\ write-up for this) but present
below some features of the interface that are important to know.
\begin{enumerate}
\item Unlike \fortran, \cxx\ code is case-sensitive. We have therefore
  adopted the convention that the wrappers will have
  the same name (and argument list) as their \fortran\ counterparts,
  but written in lower case. Furthermore, to avoid possible
  name-conflicts with other codes, all the wrappers are assigned to the
  name\-space~\xtt{QCDNUM}. We thus have
  \bcxx
  \abovedisplayskip=0pt
  \begin{equation}\nonumber
  \text{\xtt{call SUB(arguments)}} \qquad \rar \qquad
  \text{\xtt{QCDNUM::sub(arguments);}}
  \end{equation}
  \ecxx
\item In the \fortran\ description of the \qcdnum\ routines we use
  the implicit type declaration that, unless otherwise stated,  all
  variables are in \xtt{double precision} except those with names
  starting with the letter
  \xtt{I}--\xtt{N} (or \xtt{i}--\xtt{n}) which are of type~\xtt{integer}. In
  \cxx\ each variable must be explicitly typed, for example,
  \bcxx
  \abovedisplayskip=0pt
  \begin{equation}\nonumber
  \begin{array}{l}
  \text{\xtt{implicit double precision (a-h,o-z)}} \\
  \text{\xtt{ix = IXFRMX(x)}}
  \end{array}
  \quad \rar \quad
  \begin{array}{l}
  \text{\xtt{double x;}} \\
  \text{\xtt{int ix = QCDNUM::ixfrmx(x);}}
  \end{array} 
  \end{equation}
  \ecxx
\item The wrapper for logical functions returns an \xtt{int} and not a
  \xtt{bool}.\footnote{In \cxx\ any nonzero (zero) value evaluates
  as true (false) in logical expressions.}
  \bcxx
  \abovedisplayskip=0pt
  \begin{equation}\nonumber
  \begin{array}{l}
  \text{\xtt{logical gridpoint}} \\
  \text{\xtt{gridpoint = XXATIX(x,ix)}} \\
  \text{\xtt{if(gridpoint) then ...}}
  \end{array}
  \quad \rar \quad
  \begin{array}{l}
  \text{\xtt{double x; int ix; }} \\
  \text{\xtt{int gridpoint = QCDNUM::xxatix(x,ix); }} \\
  \text{\xtt{if(gridpoint) \{ ... }}
  \end{array} 
  \end{equation}
  \ecxx
\item The type of character input arguments should be \xtt{string}.
  String literals are delimited by double quotes in \cxx\
  and by single quotes in standard \fortranzz.
  \bcxx
  \abovedisplayskip=0pt
  \begin{equation}\nonumber
  \begin{array}{l}
  \text{\xtt{character*50 file}} \\
  \text{\xtt{file = 'example.log'}} \\
  \text{\xtt{call QCINIT(20,file)}} \\
  \text{\xtt{call SETVAL('Alim',5.0D0)}}
  \end{array}
  \quad \rar \quad
  \begin{array}{l}
  \text{\xtt{string file = "example.log"; }} \\
  \text{\xtt{QCDNUM::qcinit(20,file); }} \\
  \text{\xtt{QCDNUM::setval("Alim",5); }}
  \end{array} 
  \end{equation}
  \ecxx
\item In \fortran\ an array index starts at one, unless
  the index range is specified as in~\xtt{pdf(-6:6)}, for example. 
  Here the gluon has index \xtt{0}
  and the (anti)quarks have a (negative) positive index according
  to the PDG~convention~\cite{ref:PDGindices}. However, this is not
  possible in \cxx\ where arrays always start with index zero.
  Thus one should account for index shifts between the \fortran\ and
  \cxx\ arrays as is shown below.
  \bcxx
  \abovedisplayskip=0pt
  \begin{equation}\nonumber
  \begin{array}{l}
  \text{\xtt{dimension pdf(-6:6)}} \\
  \text{\xtt{call ALLFXQ(1,x,q,pdf,0,1)}}\\
  \text{\xtt{gluon = pdf(0)}}
  \end{array}
  \qquad \rar \qquad
  \begin{array}{l}
  \text{\xtt{double x, q, pdf[13];}} \\
  \text{\xtt{QCDNUM::allfxq(1,x,q,pdf,0,1);}}\\
  \text{\xtt{double gluon = pdf[6];}}
  \end{array} 
  \end{equation}
  \ecxx
  \item Two-dimensional arrays in \fortran\ become one-dimensional
  arrays in the \cxx\ wrappers. This can best be handled
  by providing a pointer function $k(i,j)$ that maps the indices of
  a \fortran\ array~\xtt{A(n,m)} onto those of a~\cxx\ array \xtt{A[n*m]}
  such that
  $k(i+1,j) = k(i,j) + 1$, $k(i,j+1) = k(i,j) + n$ and $k(1,1) = 0$. 
  \bcxx
    \ \  \xtt{inline int kij(int i, int j, int n) \{ return i-1 + n*(j-1); \} }
  \ecxx \vspace{-4mm}
  Here is an example of how to use such a pointer.
  \bcxx
  \abovedisplayskip=0pt
  \begin{equation}\nonumber
  \begin{array}{l}
  \text{\xtt{dimension c(-6:6),x(8),q(5),f(8,5)}} \\
  \text{\xtt{call FTABLE(1,c,0,x,8,q,5,f,1)}}\\
  \text{\xtt{fij = f(i,j)}}
  \end{array}
  \hspace{-1mm} \rar \  
  \begin{array}{l}
  \text{\xtt{double c[13],x[8],q[5],f[8*5];}} \\
  \text{\xtt{QCDNUM::ftable(1,c,0,x,8,q,5,f,1);}}\\
  \text{\xtt{double fij = f[ kij(i,j,8) ];}}
  \end{array} 
  \end{equation}
  \ecxx
\item Particular care has to be taken when passing functions as arguments.
  An example is the evolution routine \xtt{evolfg} where the PDF values
  $f_i(x)$ at the input scale \msz\ are, in \fortran,
  entered via the user-defined function \xtt{func(i,x)} which should be 
  declared \xtt{external} in the calling routine. To the best of
  our understanding this can only be ported to \cxx\ if the
  corresponding function has its input arguments
  passed as pointers, as is shown for the input function \xtt{func}
  in the listing of \Fi{fig:cxxlisting}.
\end{enumerate}

This is then more or less all one needs to know about the
\cxx\ interface to \qcdnum.
 
\section{Acknowledgements}

We are grateful to the {\tt xFitter} developers team, particularly
to Ringaile~Pla\v{c}akyt\.{e}, Voica~Radescu, and Sasha~Glazov who
have encouraged the development of the interface. V.~B. is
supported by the European Research Council Starting Grant~`PDF4BSM'.


\end{document}

%% file: layout.tex
  \newlength{\dinwidth} \newlength{\dinmargin}

%% file: newcommands.tex
\newmdenv[%
    startcode={\footnotesize},%
    frametitlealignment=\center,%
    backgroundcolor=gray!10,linecolor=gray!10,%
    innerleftmargin=0.5cm,innerrightmargin=0.5cm,%
    innertopmargin=0.5\baselineskip,innerbottommargin=0.5\baselineskip,%
    splittopskip=\baselineskip,%
    ]{cxxbox}
    
\newcommand{\bcxx}{\par\begin{cxxbox}}
\newcommand{\ecxx}{\end{cxxbox}}

\newcommand{\beq}[1]{\begin{equation}\label{#1}}
\newcommand{\eeq}{\end{equation}}
\newcommand{\bea}[1]{\begin{eqnarray}\label{#1}}
\newcommand{\eea}{\end{eqnarray}}

\newcommand{\Fi}[1]{Figure~\ref{#1}}

  \def\Journal#1#2#3#4{{#1}~{\bf #2}, #3 (#4)}

  \def\EPJ{Eur.\ Phys.\ J.}

  \def\CPC{Comput.\ Phys.\ Commun.}

\newcommand{\qcdnum}{\mbox{\sc qcdnum}}
\newcommand{\Qcdnum}{\mbox{\sc Qcdnum}}

\newcommand{\fortran}{\mbox{\sc fortran}}

\newcommand{\cxx}{\mbox{C{\tt ++}}}
\newcommand{\fortranzz}{\mbox{\sc fortran}{\sc\footnotesize 77}}

\newcommand{\zmstf}{\mbox{\sc zmstf}}
\newcommand{\hqstf}{\mbox{\sc hqstf}}
\newcommand{\autotools}{\mbox{\sc autotools}}
\newcommand{\Autotools}{\mbox{\sc Autotools}}

\newcommand{\home}{\ensuremath{\sim}}

\newcommand{\ms}{\ensuremath{\mu^2}}
\newcommand{\msz}{\ensuremath{\mu^2_0}}
\newcommand{\as}{\ensuremath{\alpha_{\rm s}}}
\newcommand{\asmz}{\ensuremath{\alpha_{\rm s}(m_{\rm Z}^2)}}
\newcommand{\cbar}{\ensuremath{\bar{c}}}
\newcommand{\xtt}{\texttt}

\newcommand{\rar}{\rightarrow}

 {\begin{list}{}{
  \settowidth{\labelwidth}{#1}%
  \setlength{\leftmargin}{\labelwidth+\labelsep}%
  \setlength{\itemsep}{#2}}}
 {\end{list}}